# Probing electron beam induced transformations on a single defect level via automated scanning transmission electron microscopy

Kevin M. Roccapriore,[1] Matthew G. Boebinger,[1] Ondrej Dyck,[1] Ayana Ghosh,[2] Raymond R. Unocic,[1]
Sergei V. Kalinin,[3,a] and Maxim Ziatdinov[2,1,b]

[1] Center for Nanophase Materials Sciences, Oak Ridge National Laboratory, Oak Ridge, TN 37831, USA

[2] Computational Sciences and Engineering Division, Oak Ridge National Laboratory, Oak Ridge, TN 37831, USA

[3] Department of Materials Science and Engineering, University of Tennessee, Knoxville TN, 37916, USA

The robust approach for real-time analysis of the scanning transmission electron microscopy (STEM) data streams, based on the ensemble learning and iterative training (ELIT) of deep convolutional neural networks, is implemented on an operational microscope, enabling the exploration of the dynamics of specific atomic configurations under electron beam irradiation via an automated experiment in STEM. Combined with beam control, this approach allows studying beam effects on selected atomic groups and chemical bonds in a fully automated mode. Here, we demonstrate atomically precise engineering of single vacancy lines in transition metal dichalcogenides and the creation and identification of topological defects graphene. The ELIT-based approach opens the pathway toward the direct on-the-fly analysis of the STEM data and engendering real-time feedback schemes for probing electron beam chemistry, atomic manipulation, and atom by atom assembly.

[a] sergei2@utk.edu
[b] ziatdinovma@ornl.gov

Atomically resolved electron and scanning probe microscopies have become the mainstay of condensed matter physics, chemistry, materials science, and nanotechnology.[1–3] The atomically resolved images of crystalline materials now allow identifying the positions of atomic columns and individual atoms with sub-picometer precision.[4] This in turn allows mapping of the ferroelectric[5–8] and ferroelastic[9–11] order parameter fields and learning the associated physical parameters of a material.[12–14] Electron energy loss spectroscopy (EELS) has progressed from single-plane and single atom[15] chemical, valence, and orbital state mapping[16] towards mapping quasiparticle excitations such as excitons and plasmons with ~meV resolution. Finally, a new range of opportunities are now emerging in the context of the 4D STEM enabled by machine learning[17] and physics analysis methods, ranging from mapping the internal electric and magnetic fields[18–20] on the atomic level to 3D reconstructions of atomic structures.[21]

Beyond imaging, STEM-based techniques are beginning to enable the exploration of chemical reactions and manipulation of atomic and molecular species. Early reports on electron beam amorphization and crystallization of semiconductors have been available since the '90s.[22–28] Combined with the advances in beam control, this approach was extended to fabrication of atomically precise structures in $SrTiO_3$ and Si.[29,30] The electron beam has further been shown to induce structural changes on the atomic level.[31–39] The beam was shown to induce oxygen vacancy ordering in oxides,[40] phase transformations in sulfides,[41] and emergence of the mesoscopic order in layered $MnPS_3$.[42] These types of observations led to the proposal that the electron beam can be used to manipulate and build matter on the atomic level,[43] similar to scanning tunneling microscopy based atomic fabrication.[44]

Electron beam induced manipulation has been demonstrated in several materials classes. While atomic positioning in 3D solids has been demonstrated,[30,45,46] the vast majority of work so far has been confined to 2D materials. For example, e-beam sculpting has been performed on graphene, h-BN, and other TMDs,[47–51] the fabrication of monolayer Mo membranes from MoSe2 has been demonstrated[52] as well as the engineering of edge structures.[53] Chains just one atom wide have been formed through careful e-beam irradiation.[54–57] Demonstrations at the level of a single atom have been performed by Dyck[58] and Susi,[59–61] who have shown e-beam directed motion of Si atoms in graphene. In several subsequent publications Dyck et. al. showed that the technique for placing Si atoms in graphene[62] could be extended to many other atomic species[63–65] illustrating the general applicability. In a similar fashion Park et. al. showed C-dopant insertion in monolayer h-BN.[66]

However, these advances have been achieved primarily using manual control, with the human operator elucidating the strategy for manipulations and performing a set of human controlled operations intertwined with imaging. Automatic feedback controlled atomic fabrication has been demonstrated;[29,30] however, this approach relied on the Fourier transform of a line scan as a measure of crystallinity. In other words, it was highly specified to the problem being addressed. Here, we begin to explore ways of generalizing approaches to e-beam fabrication to enable autonomous operation enabled by atom detection.

Enabling fully autonomous atomic fabrication workflows requires developing robust solutions for several successive goals, including semantic segmentation of images towards decoding atomic positions, discovering the cause-and-effect relationships between beam position and materials response, and building upon these to discover plausible experimental pathways towards target structure.[43] Previously, the models trained on simulated data showed great potential to generalize to experimental data with manual tuning.[67,68] However, the pre-trained models were applied to the already acquired experimental data and for the most part, focused on "proving the concept", so the cost of error (misclassification) was minimal and adaptation for new data sets could be performed via e.g. transfer learning. On the other hand, for real-time applications, a model needs to make an accurate prediction in a matter of seconds (or even less) with no time for the human operator to double-check it. Unfortunately, the pre-trained models do not typically perform well in such scenarios as they can be easily thrown off by changes in imaging conditions, sample or imaging artifacts, etc.[69] This is similar to how self-driving cars perform badly under unusual weather conditions or because of camera "glitches." Previously, we demonstrated using pre-acquired data how the ensemble learning and iterative training (ELIT) approach can be used to overcome some of these limitations.[70] However, this approach hasn't yet been incorporated in a real experiment to analyze streaming data.

Here, we develop and experimentally demonstrate a robust deep ensemble learning-based approach for real time analysis of the data flow in STEM, and demonstrate automated experiments based on beam-induced atomic transformations of specific groups. In recent years, automation and automated experiments in STEM (as well as in scanning probe microscopies[71]) have been gaining traction and have consequently enabled a variety of applications, ranging from learning structure-property relationships on the fly for self-guided measurements to automated analysis guided by sparse data analytics.[72–77] Almost every automation scheme has required fast and reliable analysis of potentially multidimensional STEM data, which for the past few years has been developed largely based on deep learning methods.[78] We present here atomically precise engineering of single vacancy lines (SVLs) in transition metal dichalcogenides and creation and identification of topological defects in graphene.

**I. ELIT framework**

The key element required to enable automated STEM experiments is the development of the robust and reliable workflows for the real-time conversion of the data stream from the detector to the atomic identities and coordinates that can be further piped to the computational environment, used for feedback in beam manipulation, or atomic fabrication. While for sufficiently high signal to noise ratios this can in principle be accomplished by simple peak finding or Hough transform algorithms, they usually require considerable manual tuning and are not robust with respect to the changes in resolution or sampling. The conventional deep convolutional neural networks (DCNN) may yield performance well above human eye capabilities and work for low signal to noise ratios;[67] however, the small changes in image resolution or sampling require retraining the network for a dissimilar data set. While not a limiting factor for the off-line data analysis, it precludes real-time data analysis when the data stream dynamically evolves in response to operator actions. Note that this limitation is common for other imaging tasks, and the considerable delay of autonomous driving compared to early expectations is partially attributed to similar behaviors. Correspondingly, here we describe and implement a robust approach that enables the real-time data stream analysis in STEM.

*Training data: simulations and augmentation*

Since manually labelling all the atoms in experimental data is rarely feasible, one usually uses simulated electron microscopy images with well-defined ground truths (atomic positions) to create a training set.[79] However, the model trained only on simulations may generalize poorly to the experimental data if the forward model used in simulations is only a crude approximation of real physical processes and/or the simulations don't account for instrumental noise and other non-idealities. To improve the generalization capacity, we use carefully crafted data augmentation pipelines aimed at reducing the gap between the training (simulations) and application (experimental) domain distributions. The data augmentation includes random zoom-ins, rotations, contrast scaling, and noise. At the same time, the data augmentation does *not* involve operations that are not aimed specifically at accounting for model limitations or instrumental non-idealities.

*Deep Ensemble Learning*

As noted earlier, the robust generalization of the already trained models to the out-of-distribution (OOD) data is an ongoing challenge in contemporary deep learning precluding its deployment in many critical applications. This can be traced to the purely correlative nature of the modern deep learning techniques that do not necessarily learn the true casual mechanisms. While for STEM applications, the augmentation of the training data helps reducing the shift between training and application domain distributions, it doesn't solve the underlying problem itself. Furthermore, it is simply not possible to account in the training data for all possible irregularities and anomalies during the experiment. In the absence of techniques to fully address this challenge, one must resort to reliably identifying the OOD data samples. One of the most promising approaches is based on the (fully) Bayesian neural networks. Here, the constant weights are replaced with probabilistic distributions and the Hamiltonian Monte Carlo (HMC) sampling is used to infer the posterior distributions of weights.

The predictive mean and variance ('uncertainty') on the new set of inputs $X_*$ is then given by

$$\hat{f}_* = \frac{1}{N}\sum_{n=1}^{N} P(X_*|w^n, D), \quad \text{(Eq. 1a)}$$

$$U[f_*] = \frac{1}{N}\sum_{n=1}^{N}(f_*^n - \hat{f}_*)^2, \quad \text{(Eq. 1b)}$$

where $w^n \sim P(w|D)$ are draws from the posterior, *D* is the training data, and *N* is the total number of the HMC samples. Unfortunately, the HMC-based deep learning is computationally expensive even on modern GPU and TPU devices and is therefore not suitable for automated and autonomous experimentation in microscopy. Recently, several studies have demonstrated that deep ensembles can provide a good approximation of the posterior predictive distribution.[80–83] In this approach, one uses a stochastic gradient descent to train an ensemble of neural networks, each with a different (pseudo-)random initialization and a different shuffling of the mini-batches of training samples. The deep ensemble prediction on new data is then performed as described by Eq 1 but with $w^n$ now representing different ensemble weights.

We also note a different role of ensembles for small and large distribution shifts. For small distribution shift, the deep ensembles provide improved generalization capacity, and their uncertainty estimates can be used to judge the reliability of the predictions. For larger distribution shifts, the deep ensembles increase the probability of identifying a model producing arti-fact free predictions on new data or on a subset of new data.

*Iterative Training*

After applying an ensemble of models trained on simulated data to the experimental data, we can identify a subset of 'good' predictions (e.g., based on the uncertainty estimates or the domain expert judgment) to train a new ensemble, this time based on the experimental data.[70] This may involve several postprocessing steps such as, for example, manual re-labeling of the incorrect predictions and/or removing the systematically produced artifacts using clustering techniques. In practice during live experiments, however, limited time is available and therefore the retraining step occurs without manual re-labeling, relying on the automated removal of systematic errors (such as, for example, misidentification of the amorphous regions as atoms), while images for which it is not possible are discarded from the new training set. After the re-training, one typically uses a new ensemble or the best model from the ensemble for the remainder of the experiment, assuming there are no significant changes to the state of the microscope. The deep learning output is then used for further analysis of the system including topological graph analysis, intensity analysis (using the predicted atomic positions), and local neighborhood analysis with mixture models. In the following, we refer to this setup as ELIT.

**II. Real-time implementation of the ELIT-based workflows**

To implement the automated experiment, the entire ELIT workflow has been integrated into the Nion Swift architecture as a plugin so that a trained ensemble may be selected and used to decode the ADF-STEM images in real time. This also allows atomic and defect classes to be assigned, and finally, specific atoms or defects can be targeted with electron irradiation with the intent of fabricating atomic-scale features and devices. The ELIT workflow is depicted in **Figure**

**1** and begins as follows: a pre-trained ensemble is selected (e.g., one ensemble per material class), and a test image is acquired, which is then passed to all models in the ensemble. Within Nion Swift[84], a data item containing a stack of decoded images for each of the models in the ensemble is generated. The prediction of multiple networks is used to generate the mean ensemble prediction and uncertainty. The mean prediction is the semantically segmented image (i.e., localization and classes of atoms), and uncertainty is the measure of confidence in this prediction. As such, it serves as an indicator of out-of-distribution behaviors, new atom classes, etc. This capability to "highlight" new behaviors as regions with high uncertainty is a key advantage of the ensemble methods.

The model whose performance appears best to the operator is then chosen for subsequent predictions. Effectively, the decoding process takes the noisy detector image and transforms it into a "pure" atomic image where all atoms are identical, i.e., atomic classification is not performed at this stage. Depending on the material system, there may be multiple classes of atoms and defects that need to be declared prior to the experiment. With graphene, for instance, ideally there is only carbon present in the system, however, there can be stray Si dopants, or topological ring defects. These possibilities should be accounted for to the degree that is necessary for the intended experiment.

Classification occurs through the use of intensity thresholding if there is an obvious and consistent difference (e.g., an impurity atom), or more generally, by performing a local analysis at each coordinate. Here, a gaussian mixture model (GMM) is applied to a stack of image patches centered at each atomic site – these patches can be created from either the decoded image or the raw image. GMM performs excellently in distinguishing atoms from one another if their local bonding environment differs. In the case of graphene, a depth-first search algorithm for traversing graphs was used to identify topological defects. Since a primary motivating factor is closely related to fabrication of atomic structures, the goal is not only to decipher atomic coordinates of specific atoms or defects, but critically - to use the electron beam to create, extend, probe, and ultimately gain control of defect creation. Hence, a large variety of experiments can be imagined when equipped with the knowledge of atomic coordinates of a particular system.

These different experiments can be easily incorporated into the Nion Swift plugin, where, as a specific example in graphene, the user may decide to target with the electron beam every $n^{th}$ carbon atom in a row within the graphene lattice each for either a static duration, e.g., 500 ms, or until a condition is met. Alternatively, a direct-write process similar to electron beam lithography methods may be conducted, where the purpose of the atomic classification is to avoid specific regions, i.e., defects created during previous patterning steps. The electron beam position is seamlessly controlled by Python commands and combined with being able to access multiple detector data streams, allows this workflow to proceed without any additional hardware. In general, we found that static times were not as effective as using feedback of the ADF intensity for when to proceed to the next target site, which is also accessible using the Swift workflow. Once $n$ target sites have been irradiated, or after a designated time has passed, a second image is acquired, decoded, and classified to assess progress. Fabrication generally requires multiple imaging steps throughout the experiment since the atomic sites begin to transform to accommodate defect generation, in addition to standard sample drift effects. Naturally then, the evolution of the beam-induced phenomena can be studied at snapshots throughout the experiment, with the understanding that the imaging step is always susceptible to inadvertently creating defects. The effect of the imaging steps is reduced by limiting dose for each image acquisition to a level where model prediction remains reliable, and if possible, maximizing the exposure time of each target site, such that there are orders of magnitude in the difference in dose between these two regimes.

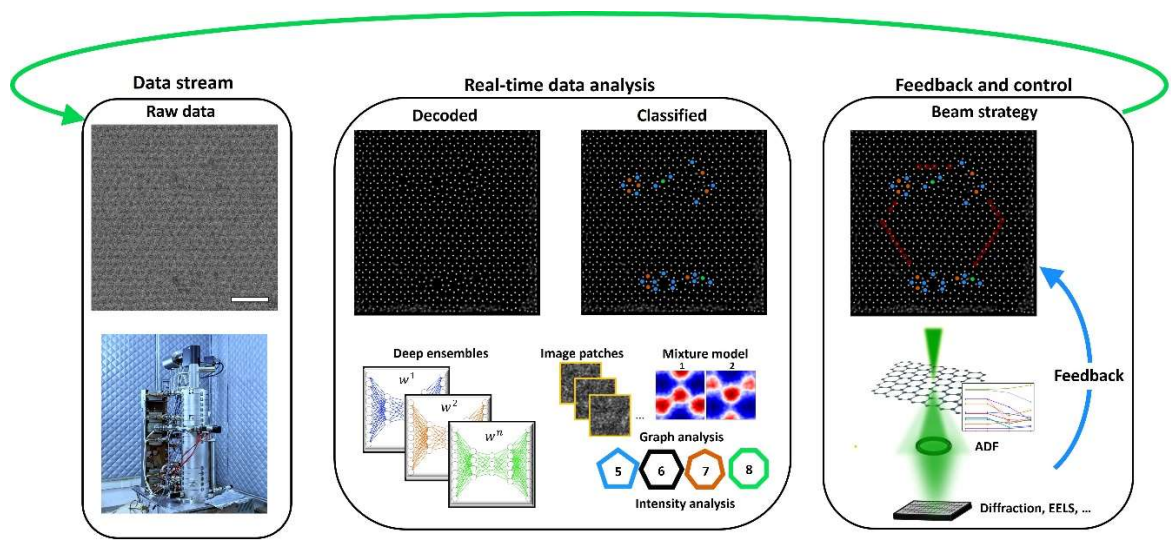

**Figure 1.** Atomic manipulation workflow. Image is acquired and passed to trained ensemble of neural networks, where raw image is decoded, and atomic coordinates are extracted. Different colors in neural networks represent different initialized weights. Image patches centered at each coordinate are generated and used for analysis of local atomic environment, where coordinates are classified using mixture models, in addition to graph analysis, intensity comparison, or a combination of all, depending on material system and possible defect evolution. According to chosen strategy, electron beam is directed to specific coordinates based on determined positions of classified atoms and defects, with possibility of using detector feedback to guide beam conditions such as dwell time. A new image is acquired once the selected beam irradiation strategy has completed, and the scheme repeats.

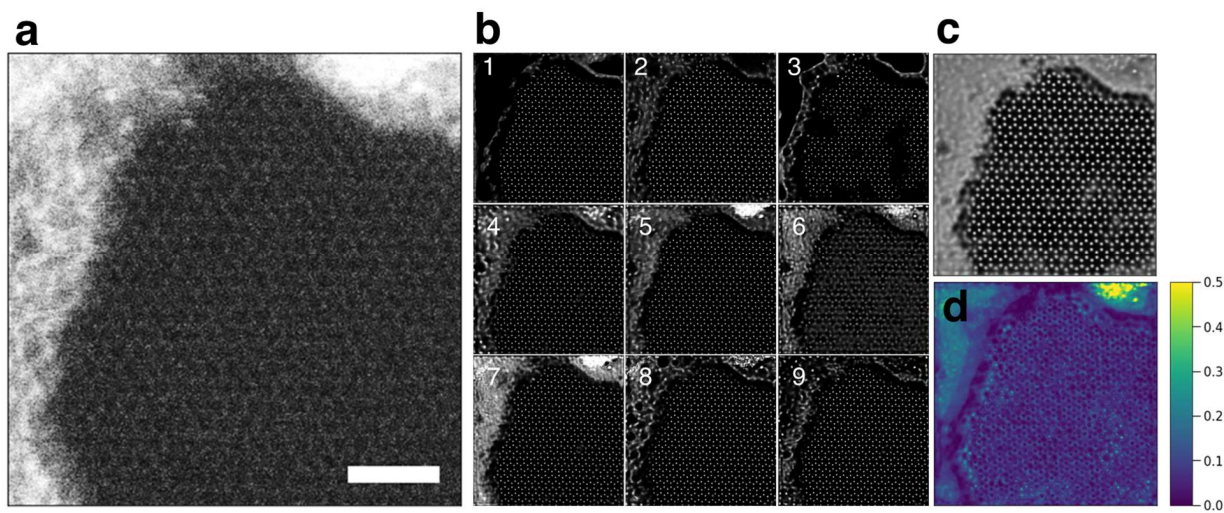

**Figure 2**. Ensemble prediction of live experimental data. Graphene lattice shown in MAADF-STEM image in (a), where first nine ensemble predictions are shown in (b). Mean ensemble prediction and corresponding relative uncertainty shown in (c) and (d) respectively. Scale bar 1 nm.

Initial application of the deep ensembles during a live experiment is evident in **Figure 2**, where we show a moderately aberrated medium angle annular dark field (MAADF)-STEM image of graphene intentionally surrounded by contamination. Graphene is notoriously difficult for obtaining high fidelity images, mainly due to the atomic weight of carbon and presence of contamination. Despite even the presence of aberrations, most of the ensemble models very accurately predict the atomic lattice, and if the mean ensemble prediction is considered, a degree of uncertainty associated with each pixel in the prediction is also provided.

If the ensemble model is suitably trained, the ensemble prediction may provide a better coordinate extraction, as shown in Figure 2(c). To obtain the ensemble prediction, the image data is required to pass through *all* models, which depending on the number of models, can last too long for practical atomic targeting. If a powerful GPU or workstation containing multiple GPUs are available, this will drastically reduce prediction times to the order of a few seconds and becomes feasible in this limit. In a general situation, we find it is more than suitable to select the single best performing model – found at the start of the microscope session - and continue to use this for the remainder of the experiment.

**III. Exploring single-atom beam reactions in graphene**

Real time implementation of the ELIT DCNN combined with beam position control enables direct exploration of the beam-induced phenomena. In these experiments, the material is imaged using low dose imaging, where this image is decoded, atoms are identified and classified, and atomic coordinates are derived. The electron beam is then positioned on predefined atomic group or groups, and the resultant changes in atomic structure are visualized. Until now, this has been preponderantly accomplished via manual control or simplified atom finding methods, while highly robust control has been elusive.

Here, we use the ELIT workflow to engender robust electron beam-induced transformation experiments. The region of interest (ROI) must be periodically sampled to retrieve the current state of the system by capturing a low-dose image followed by the return of atomic coordinates as soon as possible. The beam can then be positioned at selected coordinates (alternatively, scanned in a small window around a coordinate) where it will move to the next coordinate after a specified condition is met according to the live feedback, e.g., an intensity threshold. In the initial experiments, feedback is in the form of the integrated intensity from the ADF detector, however, any available signal can be used if it can be integrated into the workflow – e.g., energy losses, (convergent beam) diffraction patterns, optical signals, etc. The largest roadblock in this effort up until this point has been a reliable method to extract atomic coordinates during a live setting. ELIT removes this obstacle and allows to target specific atoms (or avoid certain sites) with the electron beam paving the way for atomic fabrication at the single defect level.

To demonstrate the power of utilizing real-time classification at the atomic scale, as a first example we deploy ELIT on free-standing graphene. It is known that the electron beam exposure can lead to the formation of various defects, creating, for example, the Stone-Wales and other 'topological' defects in graphene.[33,85–90] Typically, these will form even during ordinary image acquisition or scanning if operating above the knock-on energy threshold of graphene; however, the specific locations in which these defects are created are stochastic due to the probabilistic nature of the atomic removal mechanism (whether that is knock-on damage or ionization).

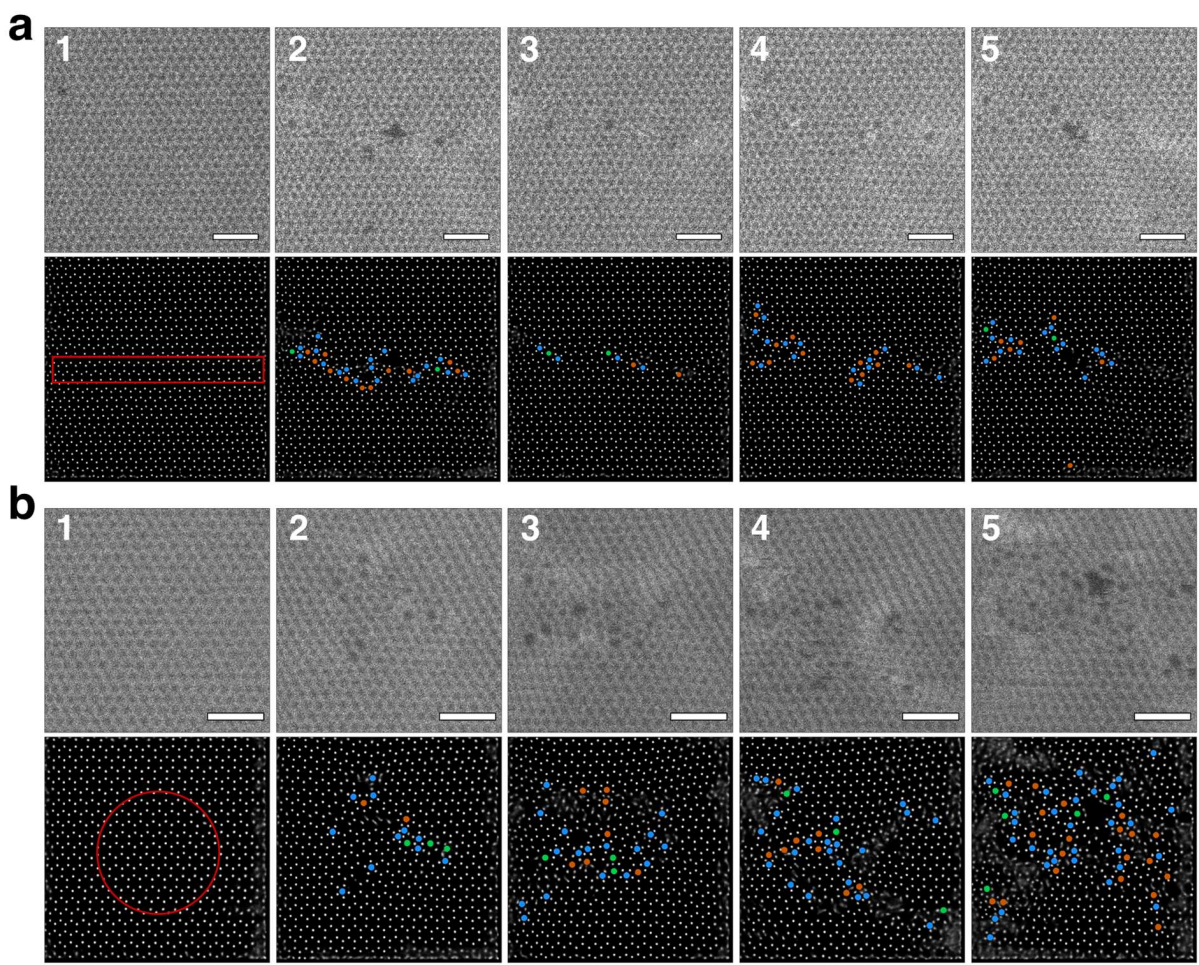

**Figure 3**. Targeted vacancy generation in graphene by atomic site-specific irradiation. Removing specific carbon atoms that form a desired shape by positioning electron beam on these atoms. MAADF-STEM images of experiments carried out at room temperature, where directly below each is the ELIT decoded and classified image. Intended defect shapes are a horizontal line (a) and a circle (b). Defect color legend: 5-ring defect, blue; 7-ring defect, orange, 8-ring defect, green. Scale bars 1 nm.

The ability to know the atomic landscape and possess control of the electron beam position allows to perform the defined experiments based on a priori defined rules. **Figure 3** illustrates two cases in single layer graphene where the goal is to generate vacancy defects in simple orientations – a line and a circle. This is not usually possible by other means – blind patterning inevitably leads to large hole formation, while routines with custom scans integrate feedback of the entire scan path, if at all. The optimum beam strategy in many cases is not obvious, and two different strategies are presented in Figures 3 and 4. This is first shown for a specimen at room temperature in Figure 3 (a) and (b), where it is noted that experiments with graphene tend to be carried out at elevated temperatures *via* a silicon heating chip to prevent contamination.[91,92] To enable operating at room temperature, the graphene was rapidly heated and cooled in-situ periodically, which we found to almost completely avoid contamination for hours at a time. Compared to hot temperatures, defect generation at room temperature was found to occur faster and more predictably as well as more localized relative to where the beam was dwelling for longer periods of time, therefore these experiments were at room temperature.

In Figure 3, both line and circle defect chains are attempted to be formed by specifically targeting pairs of carbon atoms that are expected (aided by intuition from theoretical models) to produce these chains. The frame-by-frame progression is shown, where each of the intended defect patterns are centered on the screen. With all atomic positions and defect sites known, in these examples, a rectangle or annulus is used to mask the coordinates, which are then sorted in space (left to right, and clockwise). When the beam is positioned on an atomic coordinate, a small sub-scan region a few pixels wide is centered on the coordinate, and its average intensity of the ADF detector is read out every *t* milliseconds, where *t* is generally 100 or 200 ms. The average intensity in time is then used as a criterion for when an atom has been ejected, at which point the next coordinate is targeted. This process occurs for a set amount of time or until all targets have been completed, where a fast image is acquired to assess progress. Generally, each iteration cannot proceed with a prolonged duration due to sample drift effects, therefore, several frames are collected throughout the experiment. To compensate for drift, a larger FOV image is acquired at very short acquisition times (to minimize dose and capture mesoscopic landmarks) and is used as a metric for cross correlation to determine the relative image shift. We note that while the ADF intensity in time was used, other signals may also be used to guide the process, such as electron energy loss spectroscopy (EELS) or convergent beam electron diffraction (CBED) patterns.

The line defect in Figure 3(a) is located reasonably well in the center of the field of view but is not horizontal as intended and incomplete – however, it is evident the defect chains begin to unsurprisingly form at an angle aligned with the principal axes, therefore specimen orientation relative to the scan axes is an important factor to consider. The intended circular chain of defects in Figure 3(b) does not form well using this beam exposure strategy. One positive comment here is that despite this, the defect density is rather high, hence hole formation has been prevented using this scheme. Due to such high density of defects, the graphene is observed to be strained, which is apparent in the MAADF-STEM images in Figure 3(b) images 2-5. More complete frame-by-frame videos of the targeted beam induced transformations can be found in the supplemental materials.

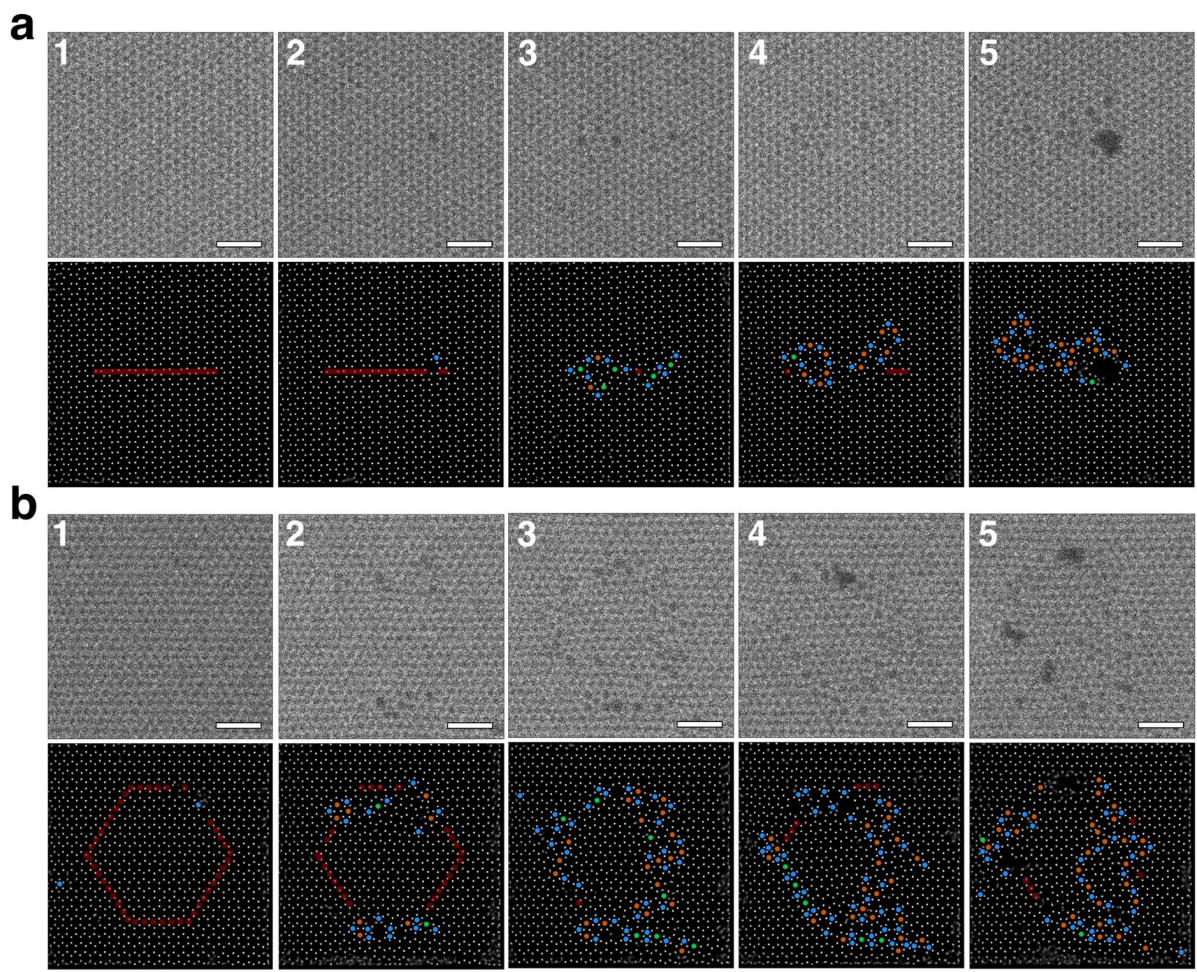

**Figure 4**. Targeted vacancy generation in graphene by dynamic avoidance patterning. Predetermined shapes are a horizontal line (a) and a hexagon (b), where the patterning positions are shown by red squares. Points of subsequent iterations are avoided if within variable distance from selected defects. Defect color legend: 5-ring defect, blue; 7-ring defect, orange, 8-ring defect, green. Scale bars 1 nm.

A difficulty with single carbon vacancies is that they are known to diffuse even close to room temperature and only become relatively stabilized in the lattice when they join with another single vacancy.[93,94] This effect is strongly enhanced by elevated temperature, which is why the previous experiment has been conducted at room temperature, but even in trying to minimize the mobility of the vacancies by operating at room temperature, the targeted defect structures were still difficult to form. The effect of beam strategies is considered, and we perform "dynamic avoidance patterning," which is introduced in **Figure 4** where similar patterns are attempted to be created. Here, instead of targeting specific pairs or clusters of atoms, the beam traces out the path of a selected pattern, much like in other direct write approaches. As before, after the first iteration of beam irradiation has been conducted, a new image is acquired, decoded, and classified. The difference now is that for the subsequent iterations of patterning, beam coordinates that reside within an "avoidance distance" to defect sites or other features where the beam should not be positioned are now removed from the beam path. Hence, this route dynamically updates the patterning coordinates based on atomic and defect classification by avoiding features that should not be irradiated. Compared to site-specific atomic targeting in graphene, avoidance patterning performs rather well, especially in the case of the hexagon defect structure. In creating a closed loop of defects in graphene as in Figure 4(b), we have formed an isolated domain bound by topological defects which, to our knowledge, has not yet been experimentally observed or intentionally created.

In all experiments in Figures 3 and 4, we note that the graphene was not destroyed. If a raster scan or beam patterning (those without atomic site-specific targeting or avoidance patterning) is performed, holes can quickly develop, and the edges of such holes are even more susceptible to beam damage and a cascading effect occurs where the material quickly breaks down. In contrast, when specific sites are targeted or avoided, the lattice is able to develop an extremely large defect density. The ELIT workflow therefore enables fabrication of non-trivial structures that cannot be organized by other means, which can allow a variety of experimentation, including both the possibility of local electronic structure modification and a degree of strain engineering in materials like graphene, and importantly without fully decomposing the structure. At the same time, more general solutions to beam strategies will ultimately require the sophistication of reinforcement learning schemes.

**IV. Single atom reactions in MoS$_2$**

As a second example, we demonstrate the ELIT workflow for MoS$_2$. In this material, electron beam irradiation can produce a variety of atomic defects. where the current focus is on subtle linear defects – so-called single vacancy lines (SVLs)[95]. SVL defects are ordered rows of single sulfur vacancies and have been observed in the STEM both thermally and by electron beam processes.[96] The motivation for placing SVLs lies in the predicted local change of the electronic band structure, in which the band gap decreases as the width of the line defect increases,[95] therefore band engineering can theoretically be performed at the atomic scale. Typically, these both will form even during ordinary image acquisition or scanning (or in the case of MoS$_2$, also with externally applied thermal energy under clean, UHV conditions[53]), however the specific locations in which these defects are created in ordinary conditions are stochastic due to the probabilistic nature of the atomic removal mechanism, making these defects of limited practical use.

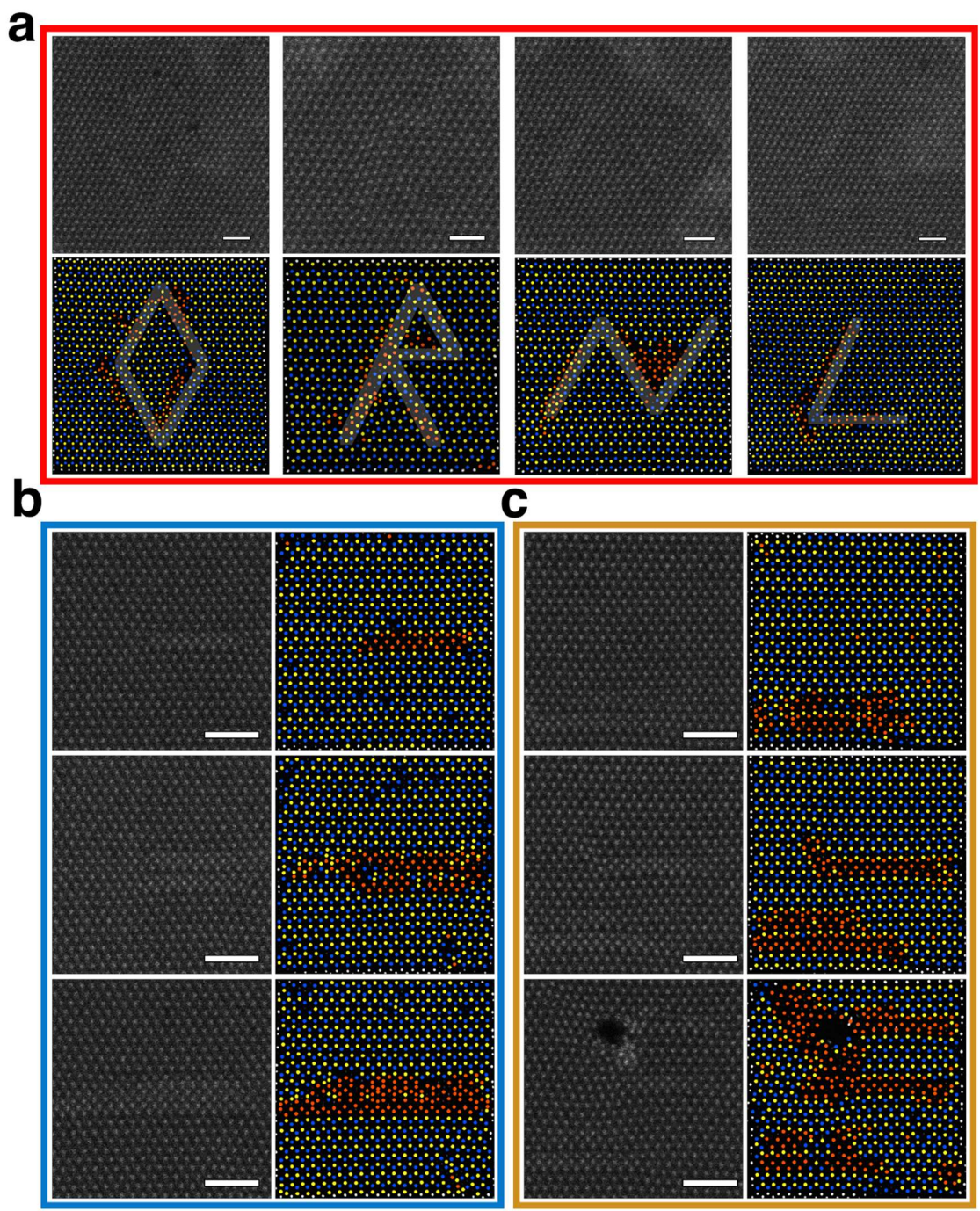

**Figure 5**. MoS$_2$ targeted defects. SVLs are placed in specific locations by the beam rather than being stochastically formed, accomplished by targeting strictly S$_2$ sites in various geometric patterns. Joining SVL segments at different angles to form laboratory logo (a), fabricating thick and multiple isolated SVLs in (b) and (c), respectively. Decoded and classified images shown
17

immediately to the right of the HAADF-STEM images. Color scheme is Mo:blue, Sulfur:yellow, Mo defected to SVL:orange. All scale bars 1 nm.

For the case of MoS$_2$, we turn to **Figure 5** which shows the defect formation of SVLs. Using ELIT with atomic site-specific beam targeting and ADF feedback, the formation of several non-trivial geometries of SVLs is shown in Figure 5. Natively, single layer MoS$_2$ consists of two atomic species – Mo and S$_2$ – and these are easily distinguished by their ADF scattering intensities. Additional defects begin to form, however, after very brief exposure to even a 60 kV electron beam, and more atomic classes may need to be understood, e.g., single sulfurs, depending on the experimental goal.

The ELIT workflow is well positioned for creation of SVL defects, since a *single* sulfur vacancy must be generated in multiple locations. To do this, the location of sulfur atoms must be known, and the method of removing sulfur atoms must be precise enough as to not remove both sulfurs. Hence, atomic identification and classification using the ELIT workflow is used to locate S$_2$ sites (separate from single sulfur sites), as well as a set of other classes based on nearest neighbor distances, intensities, or other local environmental factors. While the beam is positioned on a S$_2$ site, we found the ADF intensity averaged from a small subscan window centered on the coordinate to be an excellent method to detect when a single sulfur is removed - the intensity as a function of time can be captured and analyzed to realize when local structural changes occur. One thing to note regarding using the ADF intensity as feedback is that it is not necessarily consistent among all target atoms. This is likely because the beam is not positioned exactly at the same radial distance from the center of the atom or atomic column – so many times assigning a cutoff intensity for triggering when an atom has ejected does not reliably work and we therefore turned to using *changes* in intensity. This and other ADF feedback are shown in supplemental materials. We reiterate that this approach was not as reliable with carbon atoms in graphene, likely due to the evolution of carbon vacancies, suggesting that different systems require different modification methods. Within this framework, SVLs are also identified such that once one SVL of the correct size has been successfully fabricated, the targeting scheme can begin to work on creating the following SVL, if any remain in the experimental directive. Figure 5 emphasizes the control of the placement of SVL defects, where a variety of different geometries of SVLs were fabricated. The defects can be made along the principal crystal axes which demonstrates the level of control as in Figures 5 (a) and (c) but can also be tuned to exhibit varying thickness as in Figure 5 (b), which may in fact be a demonstrated case of band gap engineering at the atomic scale using the electron beam, however due to both delocalization effects from the surrounding pristine $MoS_2$ and potentially damaging the defects by the beam, the change in band gap is extremely challenging to experimentally verify.

**V. Summary**

It very quickly becomes evident that the atomic manipulation schemes are challenging due to a variety of reasons, but an important aspect is the fact that both the imaging tool and modification tool are the same. The general idea is to separate each mode of operation by several orders of magnitude of dose, i.e., acquire very fast images with minimal dose and target atomic sites with the maximum necessary dose. The image acquisition must of course possess enough fidelity that atomic identification is still reliably possible, but this is exactly where deep neural networks excel. Since the modification dose is set by the probabilistic nature and electron beam conditions, the only way to widen the gap further is to reduce the dose during the imaging stage. We propose that in future experiments that have access to multiple GPUs and compute stations, a defocused probe ptychographic reconstruction calculated on the fly may be performed that will lower this imaging dose by roughly an order of magnitude.

A critical point in this workflow of atomic fabrication is that while the electron beam can be precisely positioned on a specific atom or defect, the rules of atomic manipulation are not necessarily the same in all materials. This is well illustrated considering that in graphene, single vacancies can rapidly diffuse away from their point of creation, while in $MoS_2$, a single sulfur vacancy can remain relatively more fixed where it is initially formed. Moreover, using different strategies to form the desired atomic features can have a considerable impact on the outcome, as is seen in Figures 3 and 4. The atomic manipulation rules are therefore different for each material system and very likely depend on a variety of conditions – primarily incident beam energy and degree of vacuum, where the former has an effect on whether knock-on displacement or ionization occurs, and the latter can promote etching and sputtering if the UHV environment is worse than ~$10^{-8}$ Torr. In this work we developed a number of intuitive strategies that were founded in both experience and theoretical predictions but suggest also that to fully harness complete atomic fabrication, the field of reinforcement learning must be applied.

**Acknowledgements:**

This research (ELIT workflows, integration with Nion-Swift) is sponsored by the INTERSECT Initiative as part of the Laboratory Directed Research and Development Program of Oak Ridge National Laboratory, managed by UT-Battelle, LLC, for the US Department of Energy under contract DE-AC05-00OR22725. The STEM experiments were supported by the U.S. Department of Energy, Office of Science, Basic Energy Sciences, Materials Sciences and Engineering Division and Oak Ridge National Laboratory's Center for Nanophase Materials Sciences (CNMS), a U.S. Department of Energy, Office of Science User Facility.

**Data Availability**

All the deep convolutional neural networks used in the study had a Unet-type architecture and were implemented and trained via AtomAI package (https://github.com/pycroscopy/atomai). The data, the examples of data analysis, and supplementary materials are freely available at: https://github.com/kevinroccapriore/ELIT-live